\documentclass{llncs}

\usepackage{booktabs} 
\usepackage{lineno}
\usepackage{bsymb}
\usepackage{listings}

\newcommand{\ebkeyw}[1]{\textsf{#1}}

\begin{document}

\title{Engaging Millennials into Learning Formal Methods}

\author{N{\'e}stor Cata{\~n}o}

\institute{ 
\email{nestor.catano@gmail.com} 
}

\maketitle

\begin{abstract}
  This paper summarizes our experience in teaching courses on formal
  methods (FM) to Computer Science (CS) and Software Engineering (SE)
  students at various universities around the world, including
  University of Madeira (UMa) in Portugal, Pontificia Universidad
  Javeriana (PUJ) and University of Los Andes (Uniandes) in Colombia,
  Carnegie Mellon University (CMU) in the USA, and at Innopolis
  University (INNO) in the Russian Federation. We report challenges
  faced during the past 10 to 15 years to teach FM to millennials
  undergraduate and graduate students and describe how we have coped
  with those challenges. We formulate a characterization of
  millennials, based on our experience, and show how this
  characterization has shaped our decisions in terms of course
  structure and content. We show how these decisions are reflected on
  the current structure of the MSS (Models of Software Systems) course
  that currently runs as part of the MSIT-SE (Master of Science in
  Information Technology - Software Engineering) programme offered at
  INNO.  We have conducted two surveys among
  students, the first one at CMU and the second one at INNO that we
  have used to document and justify our decisions. The first survey is
  about the choice of Event-B as mathematical formalism and the second
  one is about the organization of teams of students within the classroom
  to work on software projects based on Event-B.
\end{abstract}


\section{Introduction}
\label{sec:intro}

In spite of the widespread misconception that Formal Methods (FM) are
not cost-effective (results do not outweigh the investment in time and
money), they have proven their potential to dramatically increase the
quality of software systems as conceived and developed by the IT
industry~\cite{BreunesseCHJ05,CH02,ERP:EB:15}. The IT software
industry is hence an unconquered yet alluring territory for people
working in Academia. We claim here that the best way to conquer that
territory is by properly training IT students, developing usable and
cost-effective FM tools, and addressing students' needs.

The first author taught FM related courses at UMa (University of
Madeira), in Portugal, from 2007 to 2013. He visited the
Human-Computer Interaction Institute (HCII) of Carnegie Mellon
University, in Pittsburgh, USA, in 2010, where he gave a couple of
guest lectures to students of the HCI master course offered by CMU as
part of the PUI (Programming Usable Interfaces) course.  PUI included
a Lab on Android \cite{Android} and a course project in which
students were asked to implement a \emph{usable} \cite{UsabEng}
Android app. The author then replicated PUI at UMa from 2011 to
2013. The course at UMa taught in 2011 was essentially a replica of the
course offered at CMU, however, in 2012 and 2013, the course at UMa
was modified to include the teaching of FM techniques, the modeling of
programs in Event-B, and the code generation of the core functionality
of an Android app developed during the course project. To our
knowledge, that was the first time that two, arguably, diverse topics,
FM and HCI, were combined in a single master course to formally
develop Android apps. PUI at UMa was lectured to HCI and CS students.

In the Fall of 2015, the first author joint the Faculty staff of the
CS department of INNO. He worked initially as a visiting scholar at
CMU, in Pittsburgh, PA, with the purpose of becoming a CMU certified
instructor of the Models of Software Systems (MSS) master course that
is part of the MSIT-SE (Master of Science in Information Technology -
Software Engineering) programme offered at CMU and INNO. The goal of
the MSIT-SE programme at CMU as well as at INNO is to create software
company leaders in the field of SE and to help students build
theoretical as well as practical expertise in the use of FM techniques
which they can later use in their careers. The MSS course is a FM
course taught to SE students in the Fall of every year.  It exposes
students to several FM techniques and models, including first-order
logic, state machines, concurrency, and temporal logic. The first
author started teaching an adapted version of MSS at INNO in the Fall
of 2016. This adapted version profited from our previous experience
with PUI at UMa whereby students implemented a usable and verified
Android app during their course project. The adapted version is
nurtured by the results of two surveys we have conducted. We conducted
the first survey in Pittsburgh in the Fall of 2015 among students of the
MSS course offered at CMU. The survey sheds light on the benefits of
using Event-B in the classroom. The MSS course at CMU included a
course project with 3 deliverables for the modeling and analysis of an
Infusion Pump \cite{InfPump}. The course project at INNO was
re-structured to consider the analysis and formal software development
of an Android app. The second survey was conducted among 
students of the MSS course at INNO. The goal of this second survey is
to understand how a team of  students can work together to develop
software based on Event-B. 

The second author has been teaching a model-based software development
course at PUJ (Pontificia Universidad Javeriana), Cali, Colombia, for
the past 15 years. Courses at PUJ utilize Event-B
\cite{Abrial:EB:Book:2010} as main mathematical formalism, and cover
not only techniques such as system modeling and refinement
\cite{EventB05} but also tool-assisted deductive proof of model
transformations with the Rodin platform \cite{rodin:plat}. Back in
2007, the second author invited the first one to give a couple of
guest lectures on JML \cite{JML:PrelDes:2006} (Java Modeling Language)
and Design-by-Contract \cite{ApplyingDesignContract} (DbC) for the
final part of one his undergraduate courses at the PUJ. This
constituted a breakthrough in the way we started structuring and
lecturing FM courses at our respective universities as JML and DbC
expose students to a more pragmatic approach to reasoning about
mathematical models: in addition to proving the correctness of their
models, students managed to run their models in an Object-Oriented
(O-O) programming language. We, therefore, decided to bridge
mathematical models in Event-B with JML-specified programs in Java: we
designed and implemented the EventB2Java Java code generator
\cite{sttt:codegen:15} and decided to incorporate it into the teaching
of our courses. We released the first version of EventB2Java in 2012
\cite{b2jml:tool:16}.

This paper summarizes our experience in teaching FM to students of
various universities around the world during the past decade and
more, and this experience is con. Our students have traditionally been Computer Science (CS),
Software Engineering (SE), but also Human-Computer Interaction (HCI)
students. The remainder of this paper is structured as follows. Section
\ref{sec:back} gives some basic background on the mathematical
fundamentals that are taught in our FM courses. This section may be
skipped by readers with a background in FM. Section \ref{sec:mil} shows our
characterization of millennials. This characterization has been gathered
from our own experience and from discussions with peers and
students. Section \ref{sec:inno} explains how this characterization has
been used to shape the MSS master course that is currently lectured at
INNO. It also shows the results and analysis of our first survey. It
gives a series of recommendations for restructuring the MSS
course. Section \ref{survey2} presents the results of our second
survey. It sheds light on how to conduct formal software development of
Android apps with Event-B and how development teams of students can be
organized in the classroom. Section \ref{sec:con} presents our
conclusions and discusses future work.

\section{Preliminaries on Formal Methods}
\label{sec:back}

This section provides a broad view of software development with
Event-B, JML and DbC, the EventB2Java tool and the Rodin platform. It
provides the reader with a basic understanding of the mathematical
fundamentals that are taught to students in our courses and the
tools that are used.

\subsection{Formal Methods}
The expression
\emph{Formal Method} refers to a \emph{direct} technique for constructing
\emph{dependable} systems. A system is dependable when evidence exists
that its benefits outweigh its risks. A direct technique is one that
focuses dependability on the system satisfying some \emph{critical}
properties, rather than on the functions or tasks it should perform. 
FMs provide ways to integrate these properties into the system design
and to mathematically prove system compliance with them.

\subsection{The Event-B Method}
\label{subs:back:eb}
Event-B is a formal modeling language for reactive systems that
allows the modeling of software and hardware
systems~\cite{Abrial:EB:Book:2010} altogether. It is based on
\emph{Action Systems} \cite{as:back:91}, a formalism describing the
behavior of a system by the atomic actions that the system carries
out.  An Action System describes the state space of a system and the
possible actions that can be executed in it. Event-B models are
composed of \emph{contexts} and \emph{machines}.  Contexts define
constants, uninterpreted sets and their properties expressed as
\ebkeyw{axioms}, while machines define variables and their properties,
and state transitions expressed as events. An event is composed of a
\emph{guard} and an \emph{action}. The guard represents conditions
that must hold in a state for the event to trigger. The actions
compute new values for state variables, thus performing observable
state transitions. In Event-B, systems are typically modeled via a sequence of
refinements.  First, an abstract machine is developed and verified to
satisfy whatever correctness and safety properties are desired.
Refinement machines are used to add more detail to the abstract
machine until the model is sufficiently concrete for hand or automated
translation to code.  Refinement Proof Obligations (POs) are discharged to
ensure that each refinement is a faithful model of the previous
machine, so that all machines satisfy the correctness properties of
the original.

\subsection{JML (the Java Modeling Language) and DbC (Design-by-Contract)}

JML \cite{JML:PrelDes:2006} is an interface specification language for
Java. It is designed for specifying the behavior of Java classes and
is included directly in Java source files using special comment
markers.  JML's type system includes all built-in Java types and
additional types representing mathematical sets, sequences, functions, 
and relations. JML expressions are a superset of Java expressions,
with the addition of notations for logical implication, existential
quantification, and for universal quantification. JML class
specifications can include {invariant} clauses (assertions that must
be satisfied in every visible state of the class), {initially} clauses
(specifying conditions that the post-state of every class constructor
must satisfy), and history constraints, which are similar to
invariants, with the additional ability to relate pre- and post-states
of a method.

JML offers support to Design-by-Contract (DbC). The basic idea behind
DbC is that classes and clients have a contract with each
other~\cite{ApplyingDesignContract}. To be able to call methods of a
class, a client must respect certain conditions. These conditions are
called \emph{preconditions}. Therefore, if the client respects the
method precondition, the class must guarantee that the
\emph{postcondition} will hold after the method call. The idea of
using preconditions and postconditions for specifying programs is not
recent, it dates back to a paper on formal verification written by
\emph{C.A.R. Hoare} in {1969}~\cite{Hoare69}.

\subsection{The EventB2Java Tool}
\label{sec:eb2java}

EventB2Java is a code generator for Event-B. It generates
JML-specified Java implementations of Event-B models. EventB2Java is
implemented as a plug-in of Rodin, an open-source Eclipse IDE that
provides a set of tools for working with Event-B models, e.g. it
provides an editor, a PO generator, and several
provers. EventB2Java generates an Eclipse project that includes the
JML-annotated Java implementation of the Event-B model and the
libraries needed to execute the Java code. 

\section{Millennials}
\label{sec:mil}

This section presents our perspective on who millennials are, and what
makes them different from other generation of students. Definitions
below come from our own experience, collected in our daily
interactions with students. We further discussed our definitions with
colleagues during a brainstorming session that took place at SECM
2017. Representatives from Academia and IT Industry participated in 
the discussions as well as a couple of millennials students attending
ICSE 2017. Millennials shared their view on the characterization
below.

\paragraph{\bf Curiosity} Millennials are curious by nature.

\paragraph{\bf Tech Savvy} Millennials naturally engage in technology
and the use of novel devices. They are not afraid of technology.

\paragraph{\bf Discovery-Driven} Millennials are often interested in
the most recent technological inventions of society. This does not
necessarily mean that they are interested in the fundamentals (to
know why it works) of that said technology but usually just in how it
works.

\paragraph{\bf Immediate Feedback} Millennials quite often ask for
immediate feedback on the activities they undertake, or feedback on
the results of their assessments. Likewise, they expect that tools
they use would give them immediate feedback. The typical case of this
in FM is to expect feedback from proof assistants (provers) on why a
proof rule cannot be applied at certain point, or why a proof tactic
cannot discharge a whole proof. They see provers as push-button
tools. They expect feedback from compilers on which line of code
is producing an error.

\paragraph{\bf Solution over Theory} Millennials sometimes show little
interest in the fundamentals of a solution. They certainly want to see
the solution of a problem, but they may struggle with being able to
generalize the solution to problems of the same kind. Solution over
Theory relates to Tech Savvy. Millennials are easily engaged in new
technology and know how to use it, but often do not know why it works.

\paragraph{\bf Active Learner} Millennials easily engage in activities
they are interested in and use technology to learn what interests
them. They frequently discover strategies through individual
experiments with a tool, for instance, when using 
provers, they might apply pruning steps of automatic proofs and
restart with different provers. 

\paragraph{\bf Easily Bored} Millennials get bored by things are not
interesting to them. Things that interest them are often related to
technology, social activities, media, and the Internet.

\paragraph{\bf Visually Focused} Millennials are interested in systems
and programs they can picture in their minds. Traditional FM courses
use toy examples to introduce topics and theories. Millennials are
often not interested in or struggle to understand those types of
problems. They often prefer to be presented examples they can
visualize in their minds, or are related to some particular
technology they are familiar with.

\paragraph{\bf Multi-Tasker} Millennials are often involved in multiple
activities at the same time. Those activities may be related to
Academia or not. 

\paragraph{\bf Individual Focus} Millennials struggle to work in
teams. This is perhaps not a unique feature of millennials. Lacking 
teamwork skills is true of students in countries the authors have
worked, namely, Portugal, Colombia, and Russia. It was true of the
students who shared their experiences with us during the discussion
sessions at SECM 2017, who studied in universities in the USA (one of
the students was originally from China). The reason students gave us
for the lack of teamwork skills is that millennials often put their
personal interests before the interests of their team.

\paragraph{\bf Socially Aware} Millennials often engage in social
activities. They care about society, animals, nature, and other people
around them. They enjoy the social media and the social apps.

\paragraph{\bf Learning from Failure} It is related to Immediate
Feedback. Millennials learn through failure and
counter-examples. They feel the need to see an example that
contradicts their theories.

\section{Formal Methods at Innopolis University}
\label{sec:inno}

Our efforts on teaching FM at INNO started the year 2016 after that
the first author visited CMU during the Fall of 2015. During his
visit, the first author received training in teaching MSS (Models of
Software Systems), a course that is part of the MSIT-SE (Master of
Science in Information Technology - Software Engineering) programme
that is offered by CMU. INNO has traditionally offered a similar
MSIT-SE programme in Russia in a common partnership with CMU. The
MSIT-SE programme at INNO is designed for SE professionals with one or
two years of work experience in software development. The programme
was originally structured and developed by CMU in Pittsburgh. Faculty
at INNO are trained at CMU on each core master course, and once back
at Innopolis are expected to replicate the same or adapted versions of
the respective courses. When the first author visited CMU in the Fall
of 2015, he has the opportunity to give some guest lectures on
predicate and first-order logic, UML and OCL, and, to introduce
Event-B into MSS for the first time. The following year the author was
appointed by INNO. In the Fall of 2016, he introduced Event-B to the
course syllabus of MSS. Initial teaching of Event-B at CMU started as
2 weekly hour and a half sessions. The teaching of Event-B at INNO was
extended from 2 to 4 sessions, and the third project deliverable was
re-oriented around the software development of an Android app whose
core functionality was modeled in Event-B. During his visit to CMU,
the author conducted a survey among students about their impressions
on the Event-B part of the course that it was introduced. The results
of this first survey are presented in Section \ref{survey1}. In
addition to that survey at CMU, we had informal chats with students
gathering their impressions about Event-B and discussing how it
can contribute to their future professional carriers.

MSS students at CMU and INNO had previous exposure to logic and
software development, typically covered by courses such as Discrete
Mathematics and Software Engineering. The MSS course at CMU consists
of 16 weekly classes and 16 weekly recitation sessions. Sessions are 2
hours and 45 minutes each. The course has homework assignments, which are issued
weekly and are due the following week. Each recitation session discusses
issues and challenges that took place during the homework assignment
of the previous week.  Students are exposed to propositional and
predicate logic, proof techniques, sets relations and functions,
sequences and induction, state machines, Z \cite{SpiveyIntro2Z},
concurrency, and linear temporal logic. The final sessions of the
course are dedicated to exposing students to the use of FM in
practice, for instance, the use of FM tools and techniques in
companies like Amazon or Microsoft. The author was given the
opportunity to introduce Event-B during two of the final practice
sessions of the CMU course where Professor David Garlan is the main
instructor. At INNO, the author modified the final part of the MSS
course structure. The Event-B sessions replaced the ``FM in Practice''
final sessions.

At CMU, the course project is about the modeling and verification of
an Infusion Pump \cite{InfPump}. The course project has 3 deliverables
on FSP (Finite State Process), Z, and FSP with temporal logic,
respectively. At INNO, we re-structured the course project to have 2
deliverables. Students initially turn in the core functionality of an
Android app in Event-B and its implementation in Java. For the second
project deliverable students turn in the interface of the app in
Android Studio, interfaced with the Java implementation.

\subsection{The Survey at CMU}
\label{survey1}

Changes introduced to the structure of the MSS course at INNO are
motivated by a survey conducted at CMU that encompasses 3 main
questions related to Event-B. Questions put forward the idea of
introducing Event-B into the course syllabus. In general,
modifications to the MSS course are subtle due to the tight interplay
of the course material: lectures are assessed through weekly homework
assignments, which are related to the course project, the midterm, and
the final exam. Hence, introducing Event-B into the course syllabus
entails to create a homework assignment for each Event-B session, to
add relevant questions on Event-B to the midterm and final exams, and
primarily to link Event-B to one or all the three project
deliverables.

Colleagues at CMU often asked us about the benefits of bringing
Event-B into the MSS course as compared to the use of Z. Thus, the
introduction of Event-B to the MSS course somehow came in
contraposition to the use of Z. The following hypotheses formalize the
concerns of my colleagues above. The survey presented in this Section
addresses those concerns and justifies the introduction of Event-B to
the MSS course at INNO in Russia\footnote{Section \ref{sec:EBvsZ}
  presents a comparison between Event-B and Z that is intended for the
  reader with a background in refinement calculus.}. We do not argue
that Event-B is a better or more expressive language or mathematical
formalism than Z.  Instead, we argue here that Event-B is more
suitable for millennials than Z as it responds better to their needs.
Event-B has a practical lien to code refinement and code generation
that is not quite present, in practice, in Z. At INNO, we have
re-oriented the third deliverable of MSS' course project to encompass
the formal software development of an Android app in which the core
functionality of the app is modeled in Event-B and its core
functionality is code generated with the
EventB2Java formal methods tool \cite{b2jml:tool:16}. \\

\textbf{Hypothesis 1:} Students can understand a program written in
Event-B more easily and accurately as compared to a program written in
Z implementing the same functionality. \\


\textbf{Hypothesis 2:} Event-B can easily be integrated and used to
validate, verify, animate, and reason about software systems we use in
industry.

\subsection{Student's Feedback}
\label{survey1:feedback}

We conducted a survey among the students of the MSS course at CMU. The
survey was anonymous and conducted online. 29 students answered the
survey. Answers were not mandatory, so some students left some answers
blank. \\

\noindent
\textbf{Question 1}: Overall, how would  you rate the Event-B sessions
of the MSS course?

\medskip

\begin{tabular}{|l|l|} \hline
Excellent & 4 \\ \hline
Good    & 18 \\ \hline
Neutral & 6 \\ \hline
Poor & 1 \\ \hline
Terrible & 0 \\ \hline
\end{tabular}

\bigskip

The results for this first question of the survey show that about 76$\%$ of
the students answered Good or Excellent, 21$\%$ answered Neutral, and
3$\%$ answered Poor or Terrible. \\

\noindent
\textbf{Question2}: What was your favorite part of the Event-B
sessions? 

\medskip

\begin{tabular}{|l|l|} \hline
They were motivated by {real examples} &  9 \\ \hline
The close link between Event-B and code & \\ 
generation and programming  languages    &  6 \\ \hline
Event-B's syntax is easy to understand  & 6\\ \hline
Event-B is tool supported &  2\\ \hline
I like it overall &  1\\ \hline
Nothing & 1 \\ \hline
Left blank &  4 \\ \hline
\end{tabular}

\medskip

79$\%$ of the answers given to this question point out to practical
aspects of Event-B. By ``real examples'' students mean a strong
connection to software systems. Students were presented a modeling
example of a social network in Event-B~\cite{matelas:10}. MSS is rather an
unusual course. It is a FM course, and hence strongly mathematically
oriented, lectured to SE students, who might or might not be as
mathematically strong as Computer Science students often are. This
fact compels us (FM instructors) to motivate and attract students by
presenting modeling and verification examples of applications they use
in life rather than demonstrating traditional Computer Science toy
examples. The examples must illustrate the lien between modeling and verification with software technology.\\


\noindent
\textbf{Question 3}: Which aspects of Event-B did you find attractive or unique
(that you do not find in other formalisms)?

\bigskip

\begin{tabular}{|l|l|} \hline
Its approach to software development (parachute) & 6  \\ \hline
Its support for code generation & 6  \\ \hline
Its tool support  & 2  \\ \hline
I do not know  & 1  \\ \hline
None  & 1  \\ \hline
Left blank  & 12  \\ \hline
It is easy to use  & 12  \\ \hline
\end{tabular}

\bigskip

48$\%$ of the answers given (the first 3 rows) point out to practical aspects of
Event-B. The first row makes reference to the fact that Event-B implements
refinement calculus techniques. As discussed in Section \ref{sec:con}, although
Z also offers refinement, in practice, it focuses more on model refinement
than on coding as Event-B does. For Event-B, we regularly use the
EventB2Java tool to demonstrate that mathematical logical models are
implementable. EventB2Java generates Java implementations of Event-B
models. \\

\noindent
{\bf Question 4.}  What would make the Event-B sessions better?

\bigskip

\begin{tabular}{|l|l|} \hline
6    & More lectures						 \\ \hline
5    & More examples, including code generation demos	 \\ \hline
2    & Putting Event-B sessions right after Z sessions						 \\ \hline
16  & Left blank					 \\ \hline
\end{tabular}

\bigskip

38$\%$ of the answers (the first 2 rows) point out to extending the
sessions on Event-B. The third row points out to having those sessions
right after the sessions on Z as notations of both languages are similar.

Overall, Question 1 tells about students' general satisfaction about
the Event-B part of the course.  Second and fourth answers to Question
2 provide support for hypothesis 2. The last answer to Question 3 and
the third answer to Question 2 give some indications about Hypothesis
1. Answers to Question 4 of the questionnaire tell that students
prefer to have the lectures on Event-B right after the ones on Z.

The two first answers to Question 2 are related to ``Visually
Focused'' (Section \ref{sec:mil}). The examples presented in class
relate to a social network. This seeks to stress the ``Social Aware''
aspect of millennials. All together we decided to write a series of
recommendations to modify the structure of
the MSS course at INNO which we present in Section \ref{sec:recom}. We
additionally map those recommendations to the description of
millennials presented in Section \ref{sec:mil}.

\subsection{Recommendations}
\label{sec:recom}

The following recommendations for improvement of the structure of the
MSS course are consequences of the hypotheses raised in Section
\ref{survey1}, the results of the survey presented in Section
\ref{survey1:feedback}, and informal discussions held with
students. We relate our recommendations with the characterization of
millennials presented in Section~\ref{sec:mil} as \emph{Related Aspects}.

\begin{enumerate}
\item Build a large battery of modeling examples, and homework
  assignments with questions and solutions. This recommendation is for
  most of the topics and chapters of the teacher guide book written by
  professors Garlan, Wing, and Celiku. Examples must be full-fledged
  modeling and verification examples of software systems. An artifact
  of this suggestion is writing the second part of the teacher guide
  book to include modeling and verification examples of software
  systems. \\ \emph{Related Aspects}: $(i.)$ ``Visually Focused'',
  examples must relate to systems students are familiar with rather
  than to programs. $(ii.)$ ``Discovery-Driven'', examples can relate
  to mobile applications, or to social networking sites.
\item Implement a hands-on approach to work during the recitation
  sessions.  Besides discussing issues related to the previous
  homework assignment, recitations should additionally expose students
  to the use of FM tools. For instance, recitation sessions on logical
  proofs can be supported with the use of the Coq proof assistant tool
  \cite{bertot2004}, rather than doing proofs manually on the
  board. In our experience SE students today do not like conducting
  logical proofs on paper.  The use of Coq (and other tools assistants
  such as PVS \cite{PvsUrl} and Isabelle~\cite{Pau94}) has several
  advantages for students. The use of proof assistants makes students
  aware of $i.)$ the fact that tool-assisted deductive proofs are
  feasible, and $ii.)$ that tool assistants check for us common
  mistakes we make during a proof that are otherwise difficult or
  impossible to detect by just human inspection. For instance, a
  common mistake that students make is to attempt to apply an
  inference rule that cannot be applied. Another common mistake we
  have noticed students make in paper-and-pencil proofs is to use a
  rule that opens an assumption, but forget to close that
  assumption. Students mistakenly think that the goal under the
  assumption is a discharged proof. Proof assistants like Coq detect
  these mistakes and provide users feedback via error messages with
  links to the source of the problem.  The proof under the assumption
  will not show as discharged. Our recommendation is not to forbid the
  use and practice of pencil-and-paper proofs among students. Students
  can conduct pencil-and-paper proofs of small theorems, to get
  familiar with the technique, but should use tool assistants for
  proving large or challenging theorems.  \\ \emph{Related Aspects}: $(i.)$
  ``Immediate Feedback'', the Coq tool provides immediate feedback on
  errors users make during a proof. $(ii.)$ ``Learning from Failure'',
  feedback provided by the Coq tool enables users to learn from their
  mistakes.
\item The second suggestion above is about using Coq, however, other proof
  assistants can do a good job too. The advantage of using Coq lies on
  the Curry-Howard isomorphism: a mathematical proof in classical
  logic without the excluded-middle rule is a program in the logic of
  the typed lambda-calculus. The consequences of this result are all
  positive. Students can run proofs as programs, for instance, in
  Objective Caml (an implementation of typed lambda calculus). If a
  lecture introduces a soundness proof of the translation performed by
  a parser, then the proof is just the program implementing the
  parser. There is no better way to motivate students to conduct
  proofs: proofs are programs that are part of software systems
  students can run.  \\ \emph{Related Aspects}: $(i.)$ ``Solution over
  Theory'', the theory about the syntax and semantics of a parser is
  carried down to animating a program that shows what the parser does.
\item Regarding the lectures on Natural and Structural induction, the
  key link between induction and programming is recursion. Recursive
  definitions require well-founded inductive proofs. If Coq is to be
  introduced into the MSS course, one can use Objective Caml (Coq's
  programming language) to write examples of recursive definitions,
  and Coq to formalize the algorithm in logic and discharge underlying
  Proof Obligations. Examples of recursive definitions may relate to
  data structures, for instance, for searching algorithms.  \\ \emph{Related
  Aspects}: $(i.)$ ``Solution over Theory'', the intrinsic aspects
  recursive definition proofs are boiled down to running programs in
  OCaml. 
\item Incorporate the teaching of Event-B to the MSS course. Event-B
  enables users $i.)$ to use a tool (Rodin~\cite{rodin:plat}) to write mathematical
  models about sets (In Event-B, relations are sets of pairs), $ii.)$
  to tool-check whether the set-based model is correct, $iii.)$ to
  conduct correctness proofs about set-based models, $iv.)$ and to
  generate Java code (via the EventB2Java tool) for students to
  animate formal models of software systems. At INNO, we have written
  Event-B models for the core functionality of various Android apps
  including a car racing game, a social event planner, and an
  inventory system called OpenBravo~\cite{ERP:EB:15}. 
\end{enumerate}

\subsection{Implementation of Recommendations at INNO}
\label{sec:imp}

Regarding recommendation 1, we have started the writing of the second
part of the MSS' teacher guide book. The first part of the (second
part of the) book introduces Event-B as the main modeling
formalism. This first part is not written as a manual on Event-B, but
the Event-B notation is introduced on-the-fly as examples require. The
book introduces the Rodin platform
(\url{http://wiki.event-b.org/index.php/Rodin\_Platform}), which
provides full support for writing Event-B models, for undertaking
underlying correctness proofs, and for generating model
implementations through the use of the EventB2Java plug-in tool. The
book uses EventB2Java to animate the software specifications. Up to
now we have worked on two software modeling examples. The first
example fully introduces a social network called Poporo and its
specification in predicate logic. The model includes an abstract
machine and four machine refinements. The second chapter is dedicated
to the modeling and verification in Event-B of a car racing game
called RoadFighter. The third chapter is about proof strategies with
Rodin. We want to build a recipe of strategies that students can reuse
when discharging proofs with Rodin. It would be a similar work to the
one presented in \cite{FreitasW14}.

Recommendations 2 and 3 are future work. Plans are to introduce Coq
into the course syllabus in the Fall of 2018. However, this would
require a lot of effort before the term starts. It requires $i.)$ to
re-structure the slides of the first part of the course (about
25$\%$), $ii.)$ to re-work the homework assignments to be based on Coq,
and $iii.)$ to adapt the course project to account for Coq.

Recommendation 4 is also future work. Let us discuss an example of how this
recommendation can be implemented in the classroom. We define a Stack in
Objective Caml that implements standard operations. We ask students to implement
a Map function that takes a function and a Stack object and applies the function
to each element of the Stack. The result is a new Stack obtained by applying the
function to each element of the original Stack. The Map function can be defined
recursively. We ask students whether their recursive definitions are correct or not,
and ask them to undertake the correctness proof formally. The Objective Caml program
can naturally be re-written in Coq, where the proof can be conducted.

Regarding recommendation 5, at INNO, we have extended the teaching of Event-B to
4 sessions of two hours 45 minutes each. Each weekly session has its
respective homework assignment on Event-B. The third deliverable of the course
project has been re-oriented to the development of an Android app
\cite{Android}, an Android car racing game called RoadFighter in the Fall of
2016. The Android app is structured following an MVC design pattern. The VC part
is based on OpenGL, the M part must initially be modeled in Event-B and then
code generated to Java using the EventB2Java tool. For the third course-project
deliverable, students must conduct 4 tasks. The first task asks students to use
ProB \cite{prob} to detect any likely deadlock or race condition in the Event-B
model. The second task asks students to define safety properties in
Event-B. The third task asks students to generate code, to animate it, and to
check if the code runs as expected. The fourth task asks students to
re-implement the interface of RoadFighter using {Usability Engineering}
techniques as advocated by Jakob Nielsen in \cite{UsabEng}.


In the next, we review the software modeling examples that we plan to add to
the teacher guide book. All the examples are part of case studies related to
papers published in international conferences and journals.

\begin{itemize}
\item Specification and verification of a multi-threaded task server
  \cite{SCP:PluralPulse:14,AVOCS:2012}. This modeling example demonstrates
  how a real-life server can be specified and verified using \emph{access
    permissions}, capabilities that can be associated with references in a
  program \cite{Boyland:Frac:2003}. The Novabase company has developed the
  server and provided access to large parts of its code. The specification
  of the server is based on existing documentation and discussions held with 
  Novabase engineers. 
\item Specification and verification of a social network in Event-B
  \cite{matelas:10}. This example is helpful for the purpose of illustrating
  topics such as $i.)$ sets, relations, and their operations $ii.)$ modeling
  of security properties (access permissions) in logic, and $iii.)$ inductive
  proofs.
\item Implementation of a social event planner in Event-B. The planner is
  coded as an Android applet. Users can schedule events (such as inviting to
  a wedding), and invite other people to the social event. The social event
  planner is built on top of the Event-B model of the previous example.
\item Specification and verification of an electronic purse for the JavaCard
  platform \cite{CatanoHuisman:02,BreunesseCatanoHuismanJacobs:05}. This
  example illustrates concepts such as formal specification of class
  invariants and Design-by-Contract.
\item Formal software development of the OpenBravo inventory system in
  Event-B \cite{ERP:EB:15}. This example illustrates concepts such as using
  sets and relations to model a relational database and UML
  modeling. We have implemented the inventory system in Android.    
\end{itemize}

\subsection{How does Event-B compare to Z?}
\label{sec:EBvsZ}
The Event-B modeling language \cite{Abrial:EB:Book:2010} is based on
predicate logic and set theory. Although Event-B shares essentially
the same modeling language for stating state properties as Z\index{Z},
Event-B and Z \cite{SpiveyIntro2Z,WoodcockBookOnZ} offer different
modeling mechanisms that are specialized in distinct mathematical
aspects. Event-B and Z are both models for state transition
systems. Event-B's language for expressing the dynamic behavior of
state machines is based on events. On the other hand, Z uses a rich
schema calculus mechanism for expressing the dynamic behavior of
models. Z Schema calculus and Event-B events coupled with model
refinement are different mechanisms. Z also offers refinement, but in
practice, Z focuses more on formal specification and Event-B focuses
more on model refinement and coding (whether manually written or
tool-generated).

Another major difference between Event-B and Z is in the undertaking
and use of invariants. In Event-B, each event definition produces
proof obligations that attest to the correctness of machine
invariants. These invariants might encode safety properties. On the
other hand, in Z, invariants are incorporated into the model
definitions, altering their meanings. They do not generate proof
obligations. Other differences between Event-B and Z are in the
notations used. Event-B does not have an explicit notation for variable
post-state as Z does. Z uses a primed variable notation to denote the
post-state of a variable. In Event-B, the use of a variable on the
right-hand side of an assignment denotes the value of the variable in
the pre-state of the event where the assignment is declared, and its
use on the left-hand side denotes its value in the post-state of the
execution of the event. Z uses a convention whereby the name of a
schema parameter ends by a question mark symbol. In Event-B,
parameters of an event are declared within an \ebkeyw{any}
symbol. Event preconditions can implicitly be encoded with the aid of
event guards, schema preconditions are encoded with schema invariants
that might make use of schema parameters.

Mapping Z specifications to Event-B can be tricky. One cannot say
that, for instance, a Z schema can directly be mapped into an Event-B
machine or an event. Event-B does not natively offer a feature similar
to Z schema calculus operators.  Schema calculus operators can
possibly be encoded with the aid of events in Event-B. In general, if
one is faced with the problem of porting a Z specification into
Event-B, one should opt to rethink the Z specification in Event-B and
write the Event-B model afresh, except for core definitions of sets,
relations, and carrier sets, which are pretty much the same in both
languages.
 

\section{The Survey at INNO}
\label{survey2}

Formal software development with Event-B follows what Jean-Raymond
Abrial calls ``the parachute strategy'' in which systems are first
considered from a very abstract and simple point of view, with broad
fundamental observations. This view usually comprises few simple
invariant properties that students can easily grasp, for instance,
defining what can reasonably be expected from the operation of such a
system. When writing a model for a software system in Event-B students
should write an abstract \emph{machine} (model) and then successively
write \textit{refinement} machines~\cite{EventB05}. For each refinement
machine Proof Obligations (POs) are to be discharged to ensure that it
is a proper refinement of the most abstract machines. Only once all the
machines are written and all the POs are discharged one can consider
the underlying system has completely been modeled. But, if an abstract
machine is modified, for instance, invariants are added to it, or some
definition is changed, then typically new POs are generated for all
the machines in the refinement chain, or existing proofs are to be
re-run. The worst scenario happens when a software requirement changes
or a new one is added on top of the existing ones as this typically
would break existing invariants. Pedagogically speaking this raises a
concern regarding the way members of a software development team
should work together and how team members can share their workload. If
team members work together in a way each member is in charge of
designing and tool-proving the correctness of a machine, then each time a member
introduces a change, the work of any team member in charge of a
refinement machine becomes invalid. In an opposite direction, one team
member can be in charge of writing the whole model, but then, at least
from a pedagogical perspective, this will diminish the Event-B
learning curve of the other team members. The parachute strategy
advocates for the Waterfall software development methodology in which
software requirements are set upfront and then the software
development process starts. In practice, this is quite difficult to
achieve, and even if it is achieved, it is often the case that actual
definitions are changed on-the-fly, for instance, when one decides to
encode a variable with a total and not with a partial function,
invalidating all the related and discharged POs.

During the Fall of 2017 at INNO, we re-engineered the course project as was
conceived initially during the training of the first author made at CMU
back in 2015. As for today, the main goal of the project is to re-implement
an Android app using the FM techniques discussed during the course. In 2017
we selected to develop WhatsApp
(\url{https://www.whatsapp.com/android/}) using predicate logic and code generation
techniques with EventB2Java. The development follows the standard MVC
(Model-View-Controller) design pattern \cite{Patterns:Gamma:95}. The VC
part of WhatsApp is fully developed in Android Studio
(\url{https://developer.android.com/studio/index.html}) using Usability
Engineering techniques as advocated by Jakob Nielsen in
\cite{UsabEng}. WhatsApp's core functionality is fully designed and
written in Event-B, using the Rodin platform \cite{rodin:plat}, and code
generated to Java with the EventB2Java tool.  The course project has 2
deliverables. The first deliverable is a Rodin project for the core
functionality of WhatsApp (the M part), and a Java implementation of it
generated with the EventB2Java tool. The second deliverable is the VC
implementation of WhatsApp, an Android Studio project, which must be
interfaced to the Java implementation generated for the first project
deliverable.

The survey presented in this section was conducted in the Fall of 2017 at INNO
to students of our MSS master course. The survey seeks to address the
\emph{Individual Focus} issue mentioned in Section \ref{sec:mil}. It attempts to
discover ways students can work together as part of a team when developing
software based on Event-B and Android. The survey was anonymous and all the 25
students of the course provided answers to all the questions.

\subsection{Student's Feedback}

\noindent
\textbf{Question 1}: What do you think would be the most suitable software
development methodology to develop WhatsApp with Event-B and Rodin?

\begin{enumerate}
\item Agile (requirements evolve, change at any time) 
\item Waterfall (requirements are stable, don't change)
\item Both combined 
\item Other? Which one?
\end{enumerate} 

\begin{tabular}{|l|l|l|} \hline
\textbf{Answer} & \textbf{$\#$} & \textbf{\%} \\ \hline
Agile        &  6 & 24\% \\ \hline
Waterfall   &  9 & 36\% \\ \hline
Both         & 8 &  32\% \\ \hline
Other       & 2 & 8\% \\ \hline
\end{tabular} \\

We initially gathered software requirements for WhatsApp from our
experience using the app; we focused on the WhatsApp's Android mobile
version and disregarded its web version. After we wrote the initial
software requirements document we proceeded to formalize the
requirements in the predicate calculus language of Event-B following
Abrial's recommendations in~\cite{Abrial:EB:Book:2010}. However,
afterwards, it was often the case we had discussions in and out the classroom
to clarify our understanding of the functionality of the app, for
instance, when two persons are chatting and one decides to delete a
previously sent content (message, picture or video), shall this
content be deleted from the sender, the receiver or anyone to whom the
content has been forwarded too? Would this functionality (to delete a
content item) be implemented differently if the person who is deleting the
content is the sender (the person who sent the content initially) or
the receiver of the content? All
these questions required profound discussions both in and out the
classroom as different implementations would break the
\emph{invariants} of the application. In short, although we all
(students and instructors) had agreed upon on the content of a
software requirements document for an application we all thought we
perfectly understood, in reality, we needed to iterate over the
requirements document and produce several releases. As instructors, at
some point, we decided not to include certain functionality of the
WhatsApp app as this would have caused to have too complex POs to
discharge.

This above text gives the reader an introduction to the first question
of the survey and the answers given by students. In a sense,
instructors thought that students could possibly follow a Waterfall
style of software development, but as we needed to evolve the software
requirements document, we needed to revisit our understanding of the
working of WhatsApp app and consult the customer (ourselves) on its
functionality. By looking at the results, students are more or less
equally fine for developing WhatsApp following Waterfall, Agile or
combining both methodologies. In practice, students needed to combine
both methodologies as requirements needed to change.

As for the last row of the results, 2 students selected Spiral as software
development methodology, which goes in the direction of a software project
in which software requirements evolve.\\

\noindent
\textbf{Question 2}: Did your team develop WhatsApp following the above-selected
methodology?

\begin{tabular}{|l|l|l|} \hline
\textbf{Answer} & \textbf{$\#$} & \textbf{\%} \\ \hline
Fully        &  1 & 4\% \\ \hline
Largely    &  10 & 40\% \\ \hline
Fairly         & 9 &  36\% \\ \hline
Scarcely       & 4 & 16\% \\ \hline
Not at all       & 1 & 4\% \\ \hline
\end{tabular} \\

According to the results, 80\% of the students (the 3 first rows)
followed a software methodology that they considered the most
suitable. We gave students complete freedom so as to choose any software 
methodology that they considered the most appropriate to develop WhatsApp.\\

\noindent
\textbf{Question 3}: If you decide to develop WhatsApp following an MVC
design pattern structure, how do you think your team should be organized to
develop the M (model) part of WhatsApp?

\begin{enumerate}
\item Software requirements are fixed in advance, and each team member
  develops one or several different machines; team meets at an early stage
  to decide who will develop what functionality and which machine; in the
  end, team meets again to glue all the machines together.
\item Only two team members would develop the complete functionality of
  WhatsApp in Event-B; the other two or three members would provide
  continuous feedback to the first two members. In short, you would engage
  in a ``pair programming'' discipline of working organized in groups of
  two members.
\item None.
\end{enumerate}

\begin{tabular}{|l|l|l|} \hline
\textbf{Answer} & \textbf{$\#$} & \textbf{\%} \\ \hline
Fixed     &  6 & 24\% \\ \hline
Paired   &  18 & 72\% \\ \hline
None     & 1 &  4\% \\ \hline
\end{tabular} 

Changes in software requirements are particularly cumbersome in
software development with Event-B since they might affect one
machine and therefore all the machines in its refinement-chain making
often most of the discharged POs invalid afterwards. Students can then decide to
split the number of machines (4 in this case) in equal shares among
students (4 or 5 students per team), working individually and
communicating decisions regularly as a team, or they can select some
of the team members to work in the Event-B formalization and the rest
of the members to work, for instance, in Android, in the visual
interface of the app. But then, students were also concerned about
learning Event-B properly as this was included in the final exam.
72\% of the students selected the last option of team work (second row
in the results table) in accordance with an Agile methodology of work in
which
requirements change constantly. \\

\noindent
\textbf{Question 4}: How difficult was for you to use Event-B to model the M
(model) part of WhatsApp?


\begin{tabular}{|l|l|l|} \hline
\textbf{Answer} & \textbf{$\#$} & \textbf{\%} \\ \hline
Very Hard     &  6 & 24\% \\ \hline
Hard            &  12 & 48\% \\ \hline
Moderate     & 7 &  28\% \\ \hline
Easy             & 0 &  0\% \\ \hline
Very easy     & 0 &  0\% \\ \hline
\end{tabular} \\

72\% students (the first 2 rows) found difficult to come up with an
implementation of the core functionality of WhatsApp. The initial
difficulty was of course to write a sound model for WhatsApp in
Event-B. Additional difficulties came from the use of the EventB2Java tool
which did not support some of the Event-B's syntax so that teams needed to
manually write the code generated by the tool. \\

\noindent
\textbf{Question 5}: How difficult was for you to extend the code generated
for the M (model) part of WhatsApp so that it can be used from the V (view)
part?

\begin{tabular}{|l|l|l|} \hline
\textbf{Answer} & \textbf{$\#$} & \textbf{\%} \\ \hline
Very Hard     &  5 & 20\% \\ \hline
Hard            &  13 & 52\% \\ \hline
Moderate     & 4 &  16\% \\ \hline
Easy             & 3 &  12\% \\ \hline
Very easy     & 0 &  0\% \\ \hline
\end{tabular} \\

Students needed to extend the core functionality of WhatsApp in the
following way. They needed $i.)$ to define the architecture of their
implementation of WhatsApp, and $ii.)$ either implement it or use an
existing platform that could handle concurrency of several users
chatting with each other in several chat-rooms. Students needed to
write some wrapping code that links the interface of the app developed
with Android Studio with the code generated by EventB2Java for the
core functionality of the app. 72\% of the students (the first 2 rows)
considered that implementing this extension was difficult, which was
expected by Course Instructors.

\noindent
\textbf{Question 6}: Given flexible time and project conditions, which approach
would you use to bridge/interface the Java code generated for the M part of
WhatsApp to the implementation of its V part?

\begin{enumerate}
\item You would write Event-B code for the extended functionality of the M part
  functionality and would generate code to Java with the EventB2Java tool that
  interfaces with the V part of WhatsApp. 
\item You would manually and directly implement the extended functionality in
  Java that interfaces with the V part of WhatsApp. 
\item Both combined.
\item None.
\end{enumerate}

\begin{tabular}{|l|l|l|} \hline
\textbf{Option} & \textbf{$\#$} & \textbf{\%} \\ \hline
Code generation     &  1 & 4\% \\ \hline
Manual Implementation            &  13 & 52\% \\ \hline
Both Combined                    & 4 &  40\% \\ \hline
None                      & 1 &  4\% \\ \hline
\end{tabular} \\

This question is about whether in addition to writing a model in Event-B and
using EventB2Java generate and implementation of it, students think it
would be worthwhile to attempt a similar approach for the interface of
WhatsApp. 52\% of the students think it is not worthwhile to attempt a
similar approach for the extended functionality, only 4$\%$ of them think
it is, and 40$\%$ think that they could attempt a combined
effort. Instructors of the course consider that it would be preferable to
write the interface manually given the complexity and size of graphical
libraries of Android.

\subsection{Related Aspects}
In what follows we discuss millennials' Related Aspects of our course-project on Android. 

\noindent
\paragraph{Tech Savvy.} Though most of our students have prior
experience in programming, only a bunch of them have prior experience
in programming with the Android platform. Hence, working on an Android project during
our course given them the opportunity to learn a new technology
while working on mathematical formalisms behind the scenes. 

\noindent
\paragraph{Immediate Feedback.} In our courses Event-B is introduced
with the aid of the Rodin IDE ~\cite{rodin:plat}, a platform that
provides support for writing models in Event-B. Rodin comes with a series of
provers that give students feedback when discharging POs
(Proof Obligations). 

\noindent
\paragraph{Visually Focused.} During the third project deliverable
students implement a visual interface of the Android app that links to
its core functionality. 

\noindent
\paragraph{Socially Aware.} Examples of social Android apps include
a social event planner (an app to invite people to gather around a
social event), WhatsApp, among others, all of which can be framed as
course projects.
\section{Formal Methods at Javeriana University}
\label{sec:puj}

xxxxx

From the year of 2001, Computer Science undergraduate studies at Javeriana University (JU) follows the  IEEE/ACM curriculum guidelines for Computer Science. In its 2013 version, this defines a Body of Knowledge for CS around eighteen areas, each divided into a collection of core and elective  tiers. Only core tiers are mandatory for compliance with the IEEE/ACM CS recommendation.  Formal methods are included as elective triers in two areas, parallel/distributed computing and software engineering. At  Javeriana those elective FM triers were integrated within the core curriculum, since the whole program was committed to the improvement of software quality production in the country. A one semester course on formal system modelling using the B language was 
 then programmed. Students were required to have previously taken a mathematical logic course were they were exposed to the translation of informal statements into the precise language of predicate logic, and to the main techniques for hand-proving a theorem in this logic. In the FM course students used the Atelier-B  platform for writing models of  systems and for ascertaining their safety properties expressed in predicate logic. 
 
 The proof of correctness of a system model with respect to its safety properties was done interactively with the help of the predicate logic provers provided by Atelier-B. The properties had first to be extracted from an informal system specification and then expressed rigorously  as formulas in predicate logic.  Students usually showed scarce motivation for undertaking this task, that they deemed ``too abstract'', and so had serious difficulties stating properties in such a way that discharging the required proofs were not  overly difficult. As a result the goal of convincing them FM is a practical system development  technique was compromised.
 
 After the characterization of millenials described in section \ref{sec:mil} FM teaching strategy at JU changed. Formation was divided into two courses. The first course (\emph{Introduction to System Modelling}) considers a  subset of Event-B for expressing simple models in the Rodin platform. The systems considered are informally presented in videos found in the web (e.g. an airplane gear system) and students discuss what are the most relevant aspects to consider. They target observations that can be represented by simple units, such as ``extended/retracted'' or ``open/closed''.  Students are only required to include in their models the most basic properties expressed as  arithmetic predicates. The student can observe the behaviour of the model using a simulator tool.  In this course, properties are not proven interactively for the model but automatically verified using an accompanying  model checking tool. This way students have immediate feedback both on the way observations change and on possible properties violations.
 
 \emph{Related Aspects}: (i) ``Visually Focused'' (ii) ``Discovery Driven'' (iii) ``Immediate feedback''
 
 The second course considers modelling and correctness proof of more complex models. In this case students must undertake the task of expressing complex properties in logic and construct a model in which those properties are proved to be verified. In order to adapt the course to the characterization of millenials, this is done stepwise by carefully choosing graded examples each illustrating a certain type of property. The examples come from systems the students are familiar with, such as social networks. Students are supplied (by using the Rodin platform) with a collection of different formal provers. By experimenting with these tools students discover when to use one or the other to discharge each proof obligation of the model. They then informally characterize the type of proof suitable to each prover. The course explores afterwards specific proving strategies that students relate to their informal discoveries. All concepts treated in this course are always illustrated by means of some tool.
 
 A relevant case is the development of correct computer programs by modelling. The formal model of the program is developed  step by step as a collection of ever more detailed models. Each one of these is verified with the tools as described before. When the last model only contains expressions and commands that students recognize as similar to those present in general-purpose programming languages, a tool is used to automatically translate the model into a Java program that is then run. Students then compare the execution of this program to the behavior they observed of the model when using the model-checking tool. 
  
  The ``socially aware'' characteristic of millenials is used for building motivation in FM. In addition to the examples chosen, this is enhanced by frequent in-class discussion of real-life software technology problems that have affected people lives. Students have to assess what sort of FM strategies introduced  in those technologies  might have avoided non desirable consequences. 
  
   \emph{Related Aspects}: (i) ``Active Learner'' (ii) ``Discovery Driven'' (iii) ``Immediate Feedback'' (iv) ``Socially Aware''.

\section{Conclusion and Future Work}
\label{sec:con}

It is evident that students today are different to students 10 or 15
years ago, and some of those differences are noticeable in the way
they learn. This paper discusses some of those differences and describes
how we have accounted for them in our teaching of FM.

Students learn better through the use of technology than by memorizing
a formula. We discussed this aspect in Section \ref{sec:imp} when we
proposed to use proof assistants (provers) to help students understand
better how to conduct deductive proofs. Besides recreating proof
sketches on paper or the whiteboard, Instructors should also use proof
assistants in the classroom for students to conduct deductive
proofs. Recursion is a topic difficult to understand in general,
nonetheless, proof assistants can be used to prove that recursive
definitions are well-founded, and the results of the Curry-Howard
isomorphism validate the fact of running the proofs as programs.

The use of the EventB2Java code generator is an asset in our
courses. Students are always positively surprised to see how
mathematical models based on predicate calculus relate to programs
written in Java (or another programming language). They love to execute
mathematical models to get a grasp on their behavior.

Motivating students through the use of modeling examples that can be
implemented as Android apps is also an asset in our course. Technology
attracts students. For the car racing game example mentioned in
Section \ref{sec:recom}, students use sets and relations to model car
lanes, obstacles, position and velocity for objects, among other
features. By generating implementations with the EventB2Java tool,
students can run their models and check if their implementations
function as expected.

Most of our future work is related to completing the guide book with a
battery modeling examples and homework assignments as described in
Section \ref{sec:recom}. Each example includes $i.)$ the core
functionality of the example written in Event-B, $ii.)$ all the POs
discharged with Rodin, $iii.)$ an implementation of the functionality
generated with the EventB2Java tool, and $iv.)$ an interface
implementation, e.g. an Android app implementation with Android
studio. Introducing Coq to the first part of the MSS course is also
future work.

For the Fall of 2018 we will continue working on an Android course
project with 2  course project deliverables based on the
experience gained in the Fall of 2017 with WhatsApp.

Finally, we would like to mention one of the difficulties that
millennials have regarding Learning from Failure. Millennials love to
learn from failure, and use counter-examples to check if something is
right or wrong, but the logical meaning they attach to
counter-examples is often wrong. If I say ``most water bottles are
made of plastic'', then a student might think it's not true because he
knows ``a water bottle made of glass'', without realizing that the
two sentences are not conflicting each other. The second sentence does
not make the first sentence invalid. To help students understand the
first sentence one would need to add some redundancy, let us say,
``most but not all the water bottles are made of plastic''.

\bibliographystyle{plain}
\bibliography{main-teach}

\begin{thebibliography}{10}

\bibitem{Abrial:EB:Book:2010}
Jean-Raymond Abrial.
\newblock {\em Modeling in {Event-B}: System and Software Design}.
\newblock Cambridge University Press, New York, NY, USA, 2010.

\bibitem{rodin:plat}
Jean-Raymond Abrial, Michael Butler, Stefan Hallerstede, Thai~Son Hoang, Farhad
  Mehta, and Laurent Voisin.
\newblock {Rodin: an open toolset for modelling and reasoning in Event-B}.
\newblock {\em Software Tools for Technology Transfer}, 12(6):447--466, 2010.

\bibitem{EventB05}
Jean-Raymond Abrial and Stefan Hallerstede.
\newblock Refinement, decomposition, and instantiation of discrete models:
  Application to {Event}-{B}.
\newblock {\em Fundam. Inf.}, 77(1-2):1--28, 2007.

\bibitem{InfPump}
David~E. Arney, Paul Jones, Insup Lee, and Arnab Ray.
\newblock Generic infusion pump hazard analysis and safety requirements version
  1.0.
\newblock Technical report, University of Pennsylvania, 2009.

\bibitem{as:back:91}
Ralph-Johan Back and Kaisa Sere.
\newblock {Stepwise Refinement of Action Systems}.
\newblock {\em {Structured Programming}}, 12:17--30, 1991.

\bibitem{bertot2004}
Yves Bertot and Pierre Cast{\'e}ran.
\newblock {\em Interactive Theorem Proving and Program Development. Coq'Art:
  The Calculus of Inductive Constructions}.
\newblock Springer, {LaBRI} Inria Futurs, 2004.

\bibitem{Boyland:Frac:2003}
John Boyland.
\newblock Checking interference with fractional permissions.
\newblock In R.~Cousot, editor, {\em Static Analysis: 10th International
  Symposium}, volume 2694 of {\em Lecture Notes in Computer Science}, pages
  55--72, Berlin, Heidelberg, New York, 2003. Springer.

\bibitem{BreunesseCHJ05}
Cees-Bart Breunesse, N{\'e}stor Cata{\~n}o, Marieke Huisman, and Bart Jacobs.
\newblock Formal methods for {Smart} {Cards}: An experience report.
\newblock {\em Science of Computer Programming}, Issues 1--3, 55:53--80, March
  2005.

\bibitem{BreunesseCatanoHuismanJacobs:05}
Cees-Bart Breunesse, N{\'e}stor Cata{\~n}o, Marieke Huisman, and Bart Jacobs.
\newblock Formal methods for smart cards: An experience report.
\newblock {\em Science of Computer Programming}, 55(1-3):53--80, March 2005.

\bibitem{SCP:PluralPulse:14}
N{\'e}stor Cata{\~n}o, Ijaz Ahmed, Radu Siminiceanu, and Jonathan Aldrich.
\newblock Lightweight verification of a multi-task threaded server.
\newblock {\em Science of Computer Programming}, 80, Part A(0):169--187, 2014.

\bibitem{CH02}
N{\'e}stor Cata{\~n}o and Marieke Huisman.
\newblock Formal specification of {Gemplus}' electronic purse case study.
\newblock In Lars-Henrik Eriksson and Peter~A. Lindsay, editors, {\em FME:
  Formal Methods Europe}, volume 2391 of {\em Lecture Notes in Computer
  Science}, pages 272--289, Copenhagen, Denmark, July 22-24 2002. Springer.

\bibitem{CatanoHuisman:02}
N{\'e}stor Cata{\~n}o and Marieke Huisman.
\newblock Formal specification of {Gemplus}' electronic purse case study.
\newblock In Lars-Henrik Eriksson and Peter~A. Lindsay, editors, {\em Formal
  Methods Europe (FME)}, volume 2391 of {\em Lecture Notes in Computer
  Science}, pages 272--289, Copenhagen, Denmark, July 22-24 2002.
  Springer-Verlag.

\bibitem{b2jml:tool:16}
N{\'e}stor Cata{\~n}o and V\'{i}ctor Rivera.
\newblock {EventB2Java}: A code generator for {Event-B}.
\newblock In {\em Nasa Formal Methods (NFM)}, volume 9690 of {\em LNCS}, pages
  166--171, Minneapolis, MN, USA, June 7--9 2016. Springer.

\bibitem{matelas:10}
N{\'e}stor Cata{\~n}o and Camilo Rueda.
\newblock Matelas: A predicate calculus common formal definition for social
  networking.
\newblock In M.~Frappier, editor, {\em Proceedings of ABZ 2010}, volume 5977 of
  {\em Lecture Notes in Computer Science}, pages 259--272, Qu{\'e}bec, Canada,
  2010. Springer.

\bibitem{ERP:EB:15}
N{\'e}stor Cata{\~n}o and Tim Wahls.
\newblock A case study on code generation of an erp system from {Event-B}.
\newblock In {\em Proceedings of the IEEE International Conference on Software
  Quality, Reliability and Security (QRS)}, pages 183--188, Vancouver, Canada,
  August 2015. IEEE Computer Society.

\bibitem{FreitasW14}
Leo Freitas and Iain Whiteside.
\newblock Proof patterns for formal methods.
\newblock In {\em {FM} 2014: Formal Methods, 19th International Symposium},
  pages 279--295, Singapore, May 12-16 2014. LNCS.

\bibitem{Patterns:Gamma:95}
Erich Gamma, Richard Helm, Ralph Johnson, and John Vlissides.
\newblock {\em Design patterns: elements of reusable object-oriented software}.
\newblock Addison-Wesley Longman Publishing Co., Inc., Boston, MA, USA, 1995.

\bibitem{Hoare69}
Tony {H}oare.
\newblock An axiomatic basis for computer programming.
\newblock {\em Communications of the {ACM}}, 12(10):576--580, 1969.

\bibitem{Android}
Google Inc.
\newblock {The Android Platform}.
\newblock \url{http://developer.android.com/design/index.html}, 2017.

\bibitem{JML:PrelDes:2006}
Gary~T. Leavens, Albert~L. Baker, and Clyde Ruby.
\newblock Preliminary design of {JML}: A behavioral interface specification
  language for {Java}.
\newblock {\em {ACM} {SIGSOFT}}, 31(3):1--38, 2006.

\bibitem{prob}
Michael Leuchel and Michael Butler.
\newblock {ProB: A Model Checker for B}.
\newblock In {\em FME 2003: Formal Methods}, LNCS 2805, pages 855--874, Pisa,
  Italy, 2003. Springer-Verlag.

\bibitem{ApplyingDesignContract}
Bertrand Meyer.
\newblock Applying ``design by contract".
\newblock {\em {Computer}}, 25(10):40--51, Oct. 1992.

\bibitem{UsabEng}
Jakob Nielsen.
\newblock {\em Usability Engineering}.
\newblock AP Professional, San Diego, CA, USA, 1993.

\bibitem{Pau94}
Tobias Nipkow, Markus Wenzel, and Lawrence~C. Paulson.
\newblock {\em Isabelle/HOL: A Proof Assistant for Higher-order Logic}.
\newblock Springer-Verlag, Berlin, Heidelberg, 2002.

\bibitem{PvsUrl}
{The {PVS} Specification and Verification System}.
\newblock \texttt{http://\-pvs.\-csl.\-sri.\-com/}, 2017.

\bibitem{sttt:codegen:15}
V\'{i}ctor Rivera, N{\'e}stor Cata{\~n}o, Tim Wahls, and Camilo Rueda.
\newblock Code generation for event-b.
\newblock {\em International Journal on Software Tools for Technology Transfer
  (STTT)}, 19:1--22, May 2015.

\bibitem{AVOCS:2012}
Radu Siminiceanu, N{\'e}stor Cata{\~n}o, and Ijaz Ahmed.
\newblock Automated verification of specifications with typestates and access
  permissions.
\newblock In G.~L{\"u}ttgen and S.~Merz, editors, {\em 12th International
  Workshop on Automated Verification of Critical Systems (AVOCS)}, volume~53,
  pages 1--15, Bamberg, Germany, September 18-20 2012. Electronic
  Communications of the EASST.

\bibitem{SpiveyIntro2Z}
J.~Michael Spivey.
\newblock An introduction to {Z} and formal specifications.
\newblock {\em Software Engineering Journal}, 4(1):40--50, 1989.

\bibitem{WoodcockBookOnZ}
Jim Woodcock and Jim Davies.
\newblock {\em {Using Z: Specification, Refinement, and Proof}}.
\newblock International Series in Computer Science. Prentice-Hall, Inc.,
  Oxford, England, 1996.

\end{thebibliography}

\end{document}